\documentclass[twocolumn, nofootinbib]{revtex4}
\begin{document}
\title{Variational problem for the Frenkel and the Bargmann-Michel-Telegdi (BMT) equations}

\author{A. A. Deriglazov} \email{alexei.deriglazov@ufjf.edu.br}
\altaffiliation{On leave of absence from Dep. Math. Phys., Tomsk Polytechnical University, Tomsk, Russia.}

\affiliation{Depto. de Matem\'atica, ICE, Universidade Federal de Juiz de Fora, MG, Brasil}

\begin{abstract}
We propose Lagrangian formulation for the particle with value of spin fixed within the classical theory. The Lagrangian
is invariant under non-abelian group of local symmetries. On this reason, all the initial spin variables turn out to be
unobservable quantities. As the gauge-invariant variables for description of spin we can take either the Frenkel tensor
or the BMT vector. Fixation of spin within the classical theory implies $O(\hbar)$\,-corrections to the corresponding
equations of motion.
\end{abstract}

\maketitle 
\section{Introduction}
Classical theories of spin are widely used (see [1-4] and references therein) in analysis of spin dynamics in various
circumstances and are known to agree with the calculations based on the Dirac theory. The spin variables of the Frenkel
and BMT theories obey the first-order equations of motion. On this reason, construction of the corresponding action
functional represents rather nontrivial problem. Various sets of auxiliary variables have been suggested and discussed
in attempts to solve the problem [5-11]. The present model is based on the recently developed construction of spin
surface \cite{aad1}. This represents an essentially unique $SO(n)$\,-invariant surface of $2n$\,-dimensional vector
space which can be parameterized by generators of $SO(n)$\,-group\footnote{More exactly, $(2n-3)$\,-dimensional spin
surface has natural structure of fiber bundle. Its base can be parameterized by $SO(n)$\,-generators.}. In \cite{aad2}
it has been demonstrated that $SO(3)$ spin surface leads to a reasonable model of non-relativistic spin. $SO(2, 3)$
spin surface implies the model of Dirac electron \cite{aad3}, and represents an example of pseudoclassical mechanics
\cite{aad4}. Here we demonstrate that $SO(1, 3)$ spin surface can be used to construct variational problem for unified
description of both the Frenkel and BMT theories of relativistic spin.

In the Frenkel theory \cite{fre} we include the three-dimensional
spin-vector $S^i$, $(S^i)^2=\frac{3\hbar^2}{4}$, into the
antisymmetric tensor $J^{\mu\nu}=-J^{\nu\mu}$. This required to
obey the constraint
\begin{eqnarray}\label{0.2}
J^{\mu\nu}u_\nu=0,
\end{eqnarray}
where $u_\nu$ represents four-velocity of the particle in proper-time parametrization. In the rest-frame, $u_\nu=(u_0,
0, 0, 0)$, this implies $J^{0i}=0$, so only three components of the Frenkel tensor survive, they are
$J^{ij}=2\epsilon^{ijk}S^k$. Besides, we can impose the covariant constraint
\begin{eqnarray}\label{0.3}
J^{\mu\nu}J_{\mu\nu}=6\hbar^2.
\end{eqnarray}
As in the rest frame $J^{\mu\nu}J_{\mu\nu}=8(S^i)^2$, this implies the right value of three-dimensional spin, as well
as the right number of spin degrees of freedom.

Frenkel tensor is equivalent to the four-vector\footnote{We use the Minkowski metric $\eta^{\mu\nu}=(-, +, +, +)$ and
the Levi-Civita symbol with $\epsilon^{0123}=+1$.} $S^\mu\equiv\frac{1}{4\sqrt{-u^2}}\epsilon^{\mu\nu\alpha\beta}u_\nu
J_{\alpha\beta}$, the latter obeys
\begin{eqnarray}\label{0.4}
S^\mu u_\mu=0.
\end{eqnarray}
This has been taken by Bargmann, Michel and Telegdi as the basic
quantity in their description of spin \cite{bmt}. In terms of the
BMT-vector, spin can be fixed fixed by the constraint\footnote{We
point out that $S^\mu$, being the Casimir operator of the Poincare
group, has fixed value for the Poincare IRREPs as well.}
\begin{eqnarray}\label{0.5}
(S^\mu)^2=\frac{1}{8}J^2-\frac{(J u)^2}{4u^2}=\frac{3\hbar^2}{4}.
\end{eqnarray}
Equations for the BMT-vector can be fixed \cite{bmt} from the requirements of relativistic covariance, the right
non-relativistic limit and from the compatibility with above mentioned constraints. Using the proper time as the
evolution parameter, they read\footnote{Our $\mu=\frac{g}{2}$ of BMT, and the charge is $e<0$.}
\begin{eqnarray}\label{0.1}
\dot S^\mu=\frac{\mu e}{mc}\left[(FS)^\mu+(SFu)u^\mu\right]-(\dot uS)u^\mu,
\end{eqnarray}
where $\mu$ stands for the anomalous magnetic moment, $m\dot
u^\mu=f^\mu$, and $f^\mu$ is four-force.

We are interested in to formulate a variational problem for the Frenkel and BMT classical spin theories. As compare
with the previous attempts [5-11], we look for the action functional which, besides of the transversality constraints
(\ref{0.2}), (\ref{0.4}), implies also the value-of-spin constraints (\ref{0.3}), (\ref{0.5}). We point out that mainly
due to the absence of variational problem, canonical quantization of the Frenkel and BMT theories is not developed to
date. We hope the present work may be a step towards this direction.

We construct the Frenkel tensor starting from angular momentum
\begin{eqnarray}\label{1.1}
J^{\mu\nu}=2(\omega^\mu\pi^\nu-\omega^\nu\pi^\mu),
\end{eqnarray}
of the spin "phase" space with the coordinates $\omega^\mu$ and
the conjugate momenta $\pi^\mu$. To achieve this, we restrict
dynamics of the basic variables on the spin surface determined by
$SO(1, 3)$\,-invariant equations
\begin{eqnarray}\label{1.2}
\pi^2=a_3, \qquad \omega^2=a_4, \qquad \omega\pi=0.
\end{eqnarray}
As $J^{\mu\nu}J_{\mu\nu}=8(\omega^2\pi^2-(\omega\pi)^2)=8a_3a_4$, an appropriate choice of the numbers $a_3$ and $a_4$
in Eq. (\ref{1.2}) fixes the value of spin. Besides, we impose the constraints\footnote{While we work with conjugate
momenta $p^\mu$ instead of four velocity $u^\mu$, our spin-vector obeys the BMT-condition (\ref{0.4}), see Eq.
(\ref{4.12}) below.}
\begin{eqnarray}\label{1.3}
p\omega=0, \qquad p\pi=0,
\end{eqnarray}
where $p^\mu$ stands for conjugate momentum to the world-line
coordinate $x^\mu$. Eqs. (\ref{1.3}) guarantee the transversality
(\ref{0.2}) of the Frenkel tensor. The set (\ref{1.2}),
(\ref{1.3}) contains one first-class constraint (see below).
Taking into account that each second-class constraint rules out
one phase-space variable, whereas each first-class constraint
rules out two variables, we have the right number of spin degrees
of freedom, $8-(4+2)=2$.

Dynamics of the position variable $x^\mu(\tau)$ is restricted by
the standard mass-shell condition
\begin{eqnarray}\label{0.6}
p^2+m^2c^2=0.
\end{eqnarray}
Our next task is to formulate the variational problem which
implies these constraints. Since they are written for the
phase-space variables, it is natural to start from construction of
an action functional in the Hamiltonian formalism.
%
%
We introduce the canonical pairs $(g_i, \pi_{gi})$, $i=1, 3, 4, 6, 7$, of auxiliary variables associated with the
constraints. Then the Hamiltonian action can be taken in the form
\begin{eqnarray}\label{1.4}
S_H=\int d\tau ~ p_\mu\dot x^\mu
+\pi_\mu\dot\omega^\mu+\pi_{gi}\dot g_i-H,
\end{eqnarray}
\begin{eqnarray}\label{1.5}
H=\frac12g_1(p^2+m^2c^2)+\frac12g_3(\pi^2-a_3)+\frac12g_4(\omega^2-a_4)+ \cr
g_6(p\omega)+g_7(p\pi)+\lambda_{gi}\pi_{gi}. \qquad \qquad \qquad \quad
\end{eqnarray}
We have denoted by $\lambda_{gi}$ the Lagrangian multipliers for
the primary constraints $\pi_{gi}=0$. Variation of the action with
respect to $g_i$ implies\footnote{The equation $\omega\pi=0$
appears as the third-stage constraint, see Eq. (\ref{3.3}) below.}
the desired constraints (\ref{1.2}), (\ref{1.3}) and (\ref{0.6}).

%
%

\section{Lagrangian of a theory with quadratic constraints}
Lagrangian of a given Hamiltonian theory with constraints can be
restored within the extended Lagrangian formalism \cite{aad6}.
%
%
Our constraints (\ref{1.2}), (\ref{1.3}) and (\ref{0.6}) are
either linear or quadratic with respect to momenta. For this case,
the general formalism can be simplified as follows. Consider
mechanics with the configuration-space variables $Q^a(\tau)$,
$g_{ab}(\tau)=g_{ba}$, $h^a{}_b(\tau)$ and $k_{ab}(\tau)=k_{ba}$
and with the Lagrangian action
\begin{eqnarray}\label{2.1}
S=\int d\tau\frac12g_{ab}DQ^aDQ^b-\frac12k_{ab}Q^aQ^b-\frac12M(\tilde g, h, k).
\end{eqnarray}
We have denoted $DQ^a\equiv\dot Q^a-h^a{}_bQ^b$, and $\tilde
g^{ab}$ is the inverse matrix of $g_{ab}$ This action can be used
to produce any desired quadratic constraints of the variables $Q,
P$. Indeed, denoting the conjugate momenta as $P_a$, $\pi_g$,
$\pi_h$ and $\pi_k$, the equations for $P_a$ can be solved
\begin{eqnarray}\label{2.2}
P_a=\frac{\partial L}{\partial\dot Q^a}=g_{ab}DQ^b, ~ \Rightarrow
~ \dot Q^a=\tilde g^{ab}P_b+h^a{}_bQ^b,
\end{eqnarray}
while equations for the remaining momenta turn out to be the
primary constraints $\pi_g=\pi_h=\pi_k=0$. The Hamiltonian reads
\begin{eqnarray}\label{2.3}
H=\frac12\tilde
g^{ab}P_aP_b+P_ah^a{}_bQ^b+\frac12k_{ab}Q^aQ^b+\frac12M+ \cr
\lambda_g\pi_g+\lambda_k\pi_k+\lambda_h\pi_h. \qquad \qquad
\end{eqnarray}
Then preservation in time of the primary constraints $\pi_g$ implies the quadratic constraints $P_aP_b+\frac{\partial
M}{\partial\tilde g^{ab}}=0$, and so on.

Comparing the Hamiltonian of our interest (\ref{1.5}) with the
expression (\ref{2.3}), let us take $Q^a=(x^\mu, \omega^\nu)$,
$P_a=(p^\mu, \pi^\nu)$,
\begin{eqnarray}\label{2.4}
\tilde g^{ab}=\left(
\begin{array}{cc}
g_1 & g_7\\
g_7 & g_3
\end{array}
\right), ~ h^a{}_b=\left(
\begin{array}{cc}
0 & g_6\\
0 & 0
\end{array}
\right), ~ k_{ab}=\left(
\begin{array}{cc}
g_4 & 0\\
0 & 0
\end{array}
\right), \nonumber
\end{eqnarray}
where $g_1=g_1\eta^{\mu\nu}$ and so on.  Besides, we take the "mass" term in the form $M=g_1m^2c^2-g_3a_3-g_4a_4$. With
this choice, the equation (\ref{2.3}) turns into the desired Hamiltonian (\ref{1.5}). So the corresponding Lagrangian
action reads from (\ref{2.1}) as follows
\begin{eqnarray}\label{2.5}
S=\int d\tau\frac{1}{2A}\left[g_3(Dx)^2-2g_7(Dx\dot\omega)+g_1\dot\omega^2\right]- \cr
\frac12g_1m^2c^2+\frac12g_3a_3-\frac12g_4(\omega^2-a_4). \qquad
\end{eqnarray}
We have denoted $A=\det\tilde g=g_1g_3-g_7^2$, $Dx^\mu=\dot x^\mu-g_6\omega^\mu$.

\section{Free theory}
Equations for the canonical momenta $p^\mu$ and $\pi^\mu$ of the
theory (\ref{2.5})
\begin{eqnarray}\label{3.1}
p^\mu=\frac{g_3}{A}Dx^\mu-\frac{g_1}{A}\dot\omega^\mu, \quad \pi^\mu=-\frac{g_7}{A}Dx^\mu+\frac{g_1}{A}\dot\omega^\mu,
\end{eqnarray}
can be resolved as follows
\begin{eqnarray}\label{3.1.1}
\dot x^\mu=g_1p^\mu+g_7\pi^\mu+g_6\omega^\mu, \quad \dot\omega^\mu=g_7p^\mu+g_3\pi^\mu,
\end{eqnarray}
while equations for the remaining momenta imply the primary
constraints, $\pi_{gi}=0$. Using these equations in the expression
$p\dot x+\pi\dot\omega-L$, we immediately obtain the Hamiltonian
(\ref{1.5}). Preservation in time of the primary constraints
implies the following chains of higher-stage constraints:
\begin{eqnarray}\label{3.2}
\pi_{g1}=0 ~ \Rightarrow ~ p^2+m^2c^2=0. \qquad \qquad \qquad \qquad \qquad
\end{eqnarray}
\begin{eqnarray}\label{3.3}
\left.
\begin{array}{c}
\pi_{g3}=0,  ~ \Rightarrow ~ \pi^2-a_3=0\\
\pi_{g4}=0,  ~ \Rightarrow ~ \omega^2-a_4=0
\end{array}
\right\} \Rightarrow \pi\omega=0, \Rightarrow \qquad \qquad \cr g_4=\frac{a_3}{a_4}g_3, ~ \Rightarrow ~
\lambda_{g4}=\frac{a_3}{a_4}\lambda_{g3}. \qquad \qquad
\end{eqnarray}
\begin{eqnarray}\label{3.4}
\pi_{g7}=0 ~ \Rightarrow ~ p\pi=0, ~ \Rightarrow ~ g_6=0, ~ \Rightarrow ~ \lambda_{g6}=0. \qquad \cr \pi_{g6}=0 ~
\Rightarrow ~ p\omega=0, ~ \Rightarrow ~ g_7=0, ~ \Rightarrow ~ \lambda_{g7}=0. \qquad
\end{eqnarray}
The constraints $p^2+m^2c^2=0$ and $\pi^2-a_3+\frac{a_3}{a_4}(\omega^2-a_4)=0$ form the first-class subset. This
indicates that the action (\ref{1.4}) is invariant under the two-parametric group of local transformations. It is
composed by the standard reparametrizations as well as by the following transformations with the parameter
$\gamma(\tau)$:
\begin{eqnarray}\label{3.5}
\delta\omega^\mu=\gamma g_3\pi^\mu, \qquad \delta\pi^\mu=-\gamma g_4\omega^\mu, \qquad \cr \delta g_3=(\gamma
g_3)\dot{}, \quad \delta g_4=(\gamma g_4)\dot{}, \quad \delta g_6=\gamma g_4g_7, \cr \delta g_7=-\gamma g_3g_6, \quad
\delta\lambda_{gi}=(\delta g_i)\dot{}. \qquad \qquad
\end{eqnarray}
Note that $x^\mu$, $J^{\mu\nu}$ and
$S^\mu=\frac12\epsilon^{\mu\nu\alpha\beta}p_\nu J_{\alpha\beta}$
are $\gamma$\,-invariant quantities.

Besides the constraints, the action implies the Hamiltonian equations
\begin{eqnarray}\label{3.6}
\dot x^\mu=g_1p^\mu, \qquad \dot p^\mu=0, \qquad \qquad \cr \dot\omega^\mu=g_3\pi^\mu, \qquad
\dot\pi^\mu=-g_3\frac{a_3}{a_4}\omega^\mu.
\end{eqnarray}
Obtaining these equations, we have used the constraints (\ref{3.3}) and (\ref{3.4}). The functions $g_1(\tau)$ and
$g_3(\tau)$ can not be determined neither with the constraints nor with the dynamical equations. It implies the
functional ambiguity in solutions to the equations of motion (\ref{3.6}): besides the integration constants, solution
depends on these arbitrary functions. The ambiguity of $x^\mu$ due to $g_1$ reflects the reparametrization invariance,
while the ambiguity of $\omega^\mu$ and $\pi^\mu$ due to $g_3$ is related with the $\gamma$\,-symmetry. According to
the general theory of singular systems \cite{dir, gt1, aad5}, the variables with ambiguous dynamics do not represent
the observable quantities. So, our next task is to find candidates for observables, which are variables with
unambiguous dynamics. Equivalently, we can look for the gauge-invariant variables. As the physical variables of the
spin-sector, we can take either the Frenkel tensor or the BMT-vector, both turn out to be $\gamma$\,-invariant
quantities. The ambiguity related with reparametrizations can be removed in the standard way: we assume that the
functions $x^\mu(\tau)$ represent the physical variables $x^i(t)$ in the parametric form. As it should be, dynamics of
the physical variables is unambiguous
\begin{eqnarray}\label{3.7}
\frac{dx^i}{dt}=c\frac{p^i}{p_0}, \quad \frac{dp^i}{dt}=0, \quad \frac{dJ^{\mu\nu}}{dt}=\frac{dS^\mu}{dt}=0.
\end{eqnarray}
According to the equations (\ref{3.2})-(\ref{3.4}), the variables obey also the desired constraints $p^2+m^2c^2=0$,
$J^{\mu\nu}p_\mu=0$, $J^2=8a_3a_4$, $S^\mu p_\mu=0$, $S^2=4m^2c^2a_3a_4$.

\section{Interaction with uniform electromagnetic field}
Let us consider the spinning particle with electric charge $e$ and the anomalous magnetic moment $\mu$. We take the
Hamiltonian of interacting theory in the form
\begin{eqnarray}\label{4.1}
H=\frac12g_1({\cal P}^2-\frac{e\mu}{2c}F_{\mu\nu}J^{\mu\nu}+m^2c^2)+\frac12g_3(\pi^2-a_3)+ \cr
\frac12g_4(\omega^2-a_4)+g_6({\cal P}\omega)+ g_7({\cal P}\pi)+\lambda_{gi}\pi_{gi}. \qquad
\end{eqnarray}
We consider uniform electromagnetic field, $F_{\mu\nu}=\partial_\mu A_\nu-\partial_\nu A_\mu=\mbox{const}$. We have
denoted ${\cal P}^\mu\equiv p^\mu-\frac{e}{c}A^\mu$. In contrast to $p^\mu$, the $U(1)$\,-invariant quantities ${\cal
P}^\mu$ have non-vanishing Poisson brackets, $\{{\cal P}^\mu, {\cal P}^\nu\}=\frac{e}{c}F^{\mu\nu}$. We point out that
the $\gamma$\,-symmetry survives in the interacting theory even for nonuniform field.

The Hamiltonian (\ref{4.1}) implies the constraints (\ref{3.3}) and the mass-shell condition
\begin{eqnarray}\label{4.2}
{\cal P}^2-\frac{e\mu}{2c}F_{\mu\nu}J^{\mu\nu}+m^2c^2=0,
\end{eqnarray}
while instead of (\ref{3.4}) we obtain
\begin{eqnarray}\label{4.3}
{\cal P}\pi=0, ~ \Rightarrow ~ g_6=g_1\frac{e(\mu-1)}{c^3M^2}(\pi F{\cal P}), ~ \Rightarrow ~ \lambda_{g6}\sim\lambda
g_1. \cr {\cal P}\omega=0, ~ \Rightarrow ~ g_7=-g_1\frac{e(\mu-1)}{c^3M^2}(\omega F{\cal P}), ~ \Rightarrow ~
\lambda_{g7}\sim\lambda g_1. ~
\end{eqnarray}
We have denoted $M^2=m^2-\frac{e(2\mu+1)}{4c^3}F_{\mu\nu}J^{\mu\nu}$. The constraints imply the useful consequence
\begin{eqnarray}\label{4.4}
g_6(\omega F{\cal P})+g_7(\pi F{\cal P})=0.
\end{eqnarray}
This equation can be used to verify that the quantities
$F_{\mu\nu}J^{\mu\nu}$, $M^2$ and ${\cal P}^2$ represent the
integrals of motion.

Hamiltonian equations for the basic variables read
\begin{eqnarray}\label{4.5}
\dot x^\mu=g_1u^\mu, \quad \dot{\cal P}^\mu=g_1\frac{e}{c}(Fu)^\mu,
\end{eqnarray}
\begin{eqnarray}\label{4.6}
\dot\omega^\mu=g_1\frac{e\mu}{c}(F\omega)^\mu+g_3\pi^\mu+g_7{\cal P}^\mu, \quad \cr
\dot\pi^\mu=g_1\frac{e\mu}{c}(F\pi)^\mu-\frac{a_3}{a_4}g_3\omega^\mu-g_6{\cal P}^\mu,
\end{eqnarray}
where the four-velocity $u^\mu$ is (see Eq. (\ref{4.3}))
\begin{eqnarray}\label{4.7}
u^\mu={\cal P}^\mu+\frac{g_7}{g_1}\pi^\mu+\frac{g_6}{g_1}\omega^\mu={\cal P}^\mu+\frac{e(\mu-1)}{2c^3M^2}(JF{\cal
P})^\mu.
\end{eqnarray}
Hence the interaction leads to modification of the Lorentz-force equation. Only for the "classical" value of anomalous
momentum, $\mu=1$, the constraints (\ref{4.3}) would be the same as in the free theory, $g_6=g_7=0$. Then the
four-velocity coincides with ${\cal P}$. When $\mu\ne 1$, the difference between $u$ and ${\cal P}$ is proportional to
$\frac{J}{c^3}\sim\frac{\hbar}{c^3}$. All the basic variables have ambiguous evolution. $x^\mu$ and ${\cal P}^\mu$ have
one-parametric ambiguity due to $g_1$ while $\omega$ and $\pi$ have two-parametric ambiguity due to $g_1$ and $g_3$.

The quantities $x^\mu$, ${\cal P}^\mu$ and the Frenkel tensor $J^{\mu\nu}$ are $\gamma$\,-invariants. Their equations
of motion form a closed system
\begin{eqnarray}\label{4.8}
\dot x^\mu=g_1\left[{\cal P}^\mu+\frac{e(\mu-1)}{2c^3M^2}(JF{\cal P})^\mu\right], \quad \dot{\cal
P}^\mu=\frac{e}{c}(F\dot x)^\mu,
\end{eqnarray}
\begin{eqnarray}\label{4.9}
\dot J^{\mu\nu}=g_1\frac{e}{c}\left[\mu F^{[\mu}{}_\alpha J^{\alpha\nu]}+\frac{\mu-1}{c^2M^2}{\cal
P}^{[\mu}J^{\nu]\alpha}(F{\cal P})_\alpha\right].
\end{eqnarray}
The remaining ambiguity due to $g_1$ presented in these equations reflects the reparametrization symmetry of the
theory. Assuming that the functions $x^\mu(\tau)$, $p^\mu(\tau)$ and $J^{\mu\nu}(\tau)$ represent the physical
variables $x^i(t)$, $p^\mu(t)$ and $J^{\mu\nu}(t)$ in the parametric form, their equations read
$\frac{dx^i}{dt}=c\frac{u^i}{u^0}$, $\frac{dp^\mu}{dt}=e\frac{(Fu)^\mu}{u^0}$, $\frac{dJ^{\mu\nu}}{dt}=c\frac{\dot
J^{\mu\nu}}{g_1u^0}$. As it should be, they have unambiguous dynamics.

Since $J^{\mu\nu}{\cal P}_\nu=0$, the Frenkel tensor is equivalent
to the BMT-vector constructed as follows:
\begin{eqnarray}\label{4.12}
S^\mu=\frac{1}{4\sqrt{-{\cal P}^2}}\epsilon^{\mu\nu\alpha\beta}{\cal P}_\nu J_{\alpha\beta}\equiv
\frac{1}{4\sqrt{-{\cal P}^2}}\epsilon^{\mu\nu\alpha\beta}u_\nu J_{\alpha\beta}.
\end{eqnarray}
So the physical dynamics can be described using $S^\mu$ instead of
$J^{\mu\nu}$. Using the identities
\begin{eqnarray}\label{4.13}
J^{\mu\nu}=-\frac{2}{\sqrt{-{\cal P}^2}}\epsilon^{\mu\nu\alpha\beta}{\cal P}_\alpha S_\beta, ~
\epsilon^{\mu\nu\alpha\beta}J_{\alpha\beta}=\frac{4}{\sqrt{-{\cal P}^2}}{\cal P}^{[\mu}S^{\nu]},
\end{eqnarray}
to represent $J^{\mu\nu}$ through $S^\mu$ in Eq. (\ref{4.8}), we
obtain the closed system of equations for $\gamma$\,-invariant
quantities
\begin{eqnarray}\label{4.14}
\dot x^\mu=g_1\left[{\cal P}^\mu+\frac{e(\mu-1)}{2c^3M^2}\epsilon^{\mu\nu\alpha\beta}(F{\cal P})_\nu{\cal P}_\alpha
S_\beta\right], \cr \dot{\cal P}^\mu=\frac{e}{c}(F\dot x)^\mu, \qquad \qquad \qquad \qquad
\end{eqnarray}
\begin{eqnarray}\label{4.15}
\dot S^\mu=g_1\frac{e\mu}{c}\left[(FS)^\mu+\frac{1}{{\cal P}^2}(SF{\cal P}){\cal P}^\mu\right]-\frac{1}{{\cal
P}^2}(\dot{\cal P}S){\cal P}^\mu. ~
\end{eqnarray}
These equations are written in an arbitrary parametrization of the world-line.

\section{Conclusion}
In this work we have specified the construction of spin surface \cite{aad1} for the case of $SO(1, 3)$\,-group. On this
base, we have constructed the Lagrangian action (\ref{2.5}) which describe the particle with fixed value of spin
interacting with uniform electromagnetic field. Due to the constraints (\ref{1.2}), (\ref{1.3}), the number of physical
degrees of freedom in the spin-sector is equal to 2, as it should be. The basic spin-space coordinates $\omega^\mu$,
$\pi^\nu$ are gauge non-invariant variables, hence they do not correspond to the observable quantities. We can take the
antisymmetric tensor (\ref{1.1}) as an observable quantity. For an appropriate choice of the parameters $a_3$, $a_4$,
this obeys both the transversality constraint (\ref{0.2}) and the value-of-spin constraint (\ref{0.3}). Its dynamics is
governed by the Frenkel-type equation (\ref{4.9}). Equivalently, we can take the vector (\ref{4.12}) as an observable
quantity. This is subject to the constraints (\ref{0.4}), (\ref{0.5}) and obeys the Bargmann-Michel-Telegdi equations
of motion (\ref{4.15}).

\section{Acknowledgments}
This work has been supported by the Brazilian foundation FAPEMIG.

\end{document}